\documentclass[
    ,final            
  ]
  {aipproc}

\layoutstyle{6x9}

\begin{document}

\title{WIMPless Dark Matter}

\classification{95.35.+d, 04.65.+e, 12.60.Jv}
\keywords      {dark matter, supersymmetry}

\author{Jason Kumar}{
  address={Department of Physics and Astronomy, University of
Hawai'i, Honolulu, HI 96822, USA}
}

\author{Jonathan L.~Feng}{
  address={Department of Physics and Astronomy, University of
California, Irvine, California 92697, USA}
}


\begin{abstract}
We describe the scenario of WIMPless dark matter.  In this scenario of
gauge-mediated supersymmetry breaking, a dark matter candidate in the
hidden sector is found to naturally have approximately the right relic
density to explain astronomical dark matter observations, but with a
wide range of possible masses.
\end{abstract}

\maketitle


\section{Introduction}

New gauge symmetries are ubiquitous features of physics beyond the
Standard Model, appearing in grand unified theories, many models of
supersymmetry (SUSY) breaking, and string theory.  But the appearance
of any new symmetry can potentially have an impact on dark matter.  If
any of the new symmetries survive at low energies (perhaps even as a
discrete symmetry), then the lightest charged particle will be stable
and will contribute to the non-baryonic matter density observed by
astronomical observation.  If the relic density of the stable particle
is close to the observed dark matter density, then the model provides
a dark matter candidate as an additional feature.

We consider WIMPless models~\cite{Feng:2008ya}, a new class of models
in which standard gauge-mediated SUSY-breaking (GMSB) models are
extended by the addition of a hidden gauge sector.  Since the hidden
sector is generic, the soft SUSY-breaking scale induced by GMSB in
this sector can be very different from the electroweak scale.  But we
will find that if a hidden particle at the hidden sector's soft mass
scale is stabilized by a remnant symmetry, then this particle will
have approximately the right relic density to be dark matter,
regardless of what its mass happens to be.  Remarkably, the relic
density for this candidate matches that observed for dark matter for
largely the same reason that WIMPs have the right relic density;
indeed, we will argue that this ``WIMPless miracle'' is really a
generalization of the WIMP miracle beyond the electroweak scale.

\section{Setup}

The setup we consider is a simple extension of the standard setup for
GMSB~\cite{Dine:1981za,Dine:1994vc}.  Typically, the GMSB setup
consists of a minimal supersymmetric Standard Model (MSSM) sector, as
well as a sector in which supersymmetry is broken.  The effects of
SUSY-breaking are then mediated to the MSSM sector via messengers
which are gauge-coupled to the MSSM.  Integrating out these messengers
generates a new scale in the effective field theory of the MSSM
sector, the soft scalar mass $m_{\rm soft}$.  Once the messengers and
other heavy matter are integrated out, the MSSM is chiral and $m_{\rm
soft}$ is the next energy scale; all matter either sits at this scale
(e.g., Higgs bosons, $W$, $Z$, and neutralinos), or is much lighter
(e.g., the photon and gluon).

\begin{figure}
  \label{sectors}
  \includegraphics[height=.2\textheight]{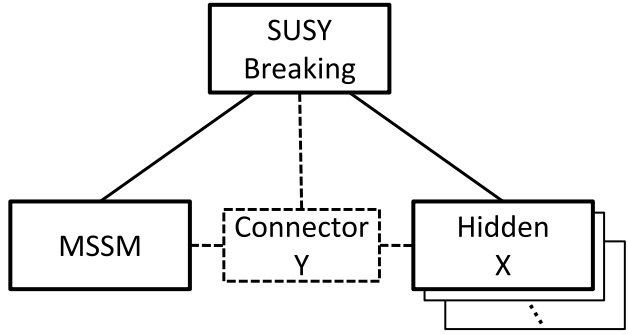}
  \caption{Sectors of the model.  SUSY breaking is mediated by gauge
interactions to the MSSM and the hidden sector(s), which contains the
dark matter particle $X$.  An optional connector sector contains
fields charged under both MSSM and hidden sector gauge groups,
which induce signals in direct and indirect searches and at colliders.
}
\end{figure}

All we add to this setup is one or more hidden sectors that are
qualitatively similar to the MSSM sector.  We only mean that these
hidden sectors are supersymmetric gauge theories that receive the
effects of SUSY-breaking from the same SUSY-breaking sector via GMSB
(see Fig.~\ref{sectors}); the gauge theory, and in particular the soft
mass scale, can be very different from the MSSM.  In the hidden
sector, once we integrate out the heavy matter, we again find that the
soft mass scale is the next energy scale, and all matter either sits
at this scale (whatever it may be), or is much lighter.  Finally, we
assume that some symmetry stabilizes a particle with mass at the
hidden sector soft mass scale.

\section{Relic Density}

In GMSB, the soft scalar mass scale is generated by a two-loop
diagram, where messengers run in the loop.  This scale is given by
\begin{equation}
\label{mmass}
m_{\rm soft} \sim \frac{g^2}{16 \pi^2} \frac{F}{M_{\rm mess}} \ ,
\end{equation}
where $g$ is the relevant gauge coupling, $F$ is the vacuum
expectation value of the SUSY-breaking $F$-term, and $M_{\rm mess}$ is
the mass of the messengers.  In particular, one expects that $F/M_{\rm
mess}$ is determined by the dynamics of the SUSY-breaking sector,
where the $F$-term is generated and where some gauge symmetries are
broken, yielding a mass scale for the messengers.  A specific example
would be the case where messengers $M_i$, $\tilde M_i$ gain mass
through the SUSY-breaking sector Yukawa coupling.  Assuming symmetries
that prevent bilinear $M_i, \tilde M_i$ couplings for the lightest
messengers, the superpotential is
\begin{equation}
W = \sum_{i= {\rm MSSM}, {\rm hidden}} \lambda_i \Phi M_i \tilde M_i \ .
\end{equation}
When SUSY breaks, the $\Phi$ field gets vacuum expectation value
$\langle \Phi \rangle = M_{\rm mess} +\theta^2 F$, and the soft mass
scale in any sector is proportional to the same quantity, $F/M_{\rm
mess}$.

We thus find that
\begin{equation}
{g_h ^4 \over m_h ^2}  \sim {g_{EW} ^4 \over m_{EW} ^2}
\propto {M_{\rm mess}^2 \over F^2} \approx {\rm const} \ ,
\end{equation}
where $g_h$ and $m_h$ are the hidden sector gauge coupling and soft
mass scale, respectively.  This is important, as the ratio $g^4/
m_{\rm soft}^2$ sets the annihilation cross-section $\sigma_{\rm ann}$
through gauge interactions for a stable particle at the soft mass
scale.  (Models where the dark matter mass is determined by loop
corrections were also studied in~\cite{Hooper:2008im}.)  Moreover, if
dark mater is thermal in the early universe, the relic density is
largely determined by $\langle \sigma_{\rm ann} v
\rangle^{-1}$~\cite{Zeldovich:1965}.

This leads to an interesting result that is the crux of WIMPless dark
matter: although the hidden sector soft mass scale could be anything,
the relic density of a stable particle at the soft scale is
essentially a universal constant, set by the physics of the
SUSY-breaking sector via the ratio $M_{\rm mess} /F$.  But one can
determine this ratio from the MSSM, and what the WIMP miracle really
shows is that this universal relic density is approximately correct to
explain the astronomical observations.  We are thus left with a good
dark matter candidate which gets the relic density right for the same
reasons as the WIMP miracle, but for the much larger mass range
$10\,{\rm MeV} < M_{DM} < 10\,{\rm TeV}$, where the lower bound is set
by the requirement that the dark matter be non-relativistic at
decoupling, and the upper bound is set by requiring perturbativity and
unitarity~\cite{Griest:1989wd}.

Indeed, we see that, from this point of view, the lack of a good dark
matter candidate in the MSSM sector is something of an accident.
Without assuming $R$-parity conservation, the two massive stable
particles of the MSSM are the electron and the LSP.  Although the
electron gets its mass from electroweak symmetry and thus might be
expected to have mass at the electroweak symmetry breaking scale
(close to the soft mass scale), in fact it is much lighter due to its
extraordinarily small Yukawa coupling.  This is basically a result of
flavor physics, which we do not understand.  And in GMSB, the LSP is
the gravitino, which is not gauge-charged and does not sit at the soft
scalar mass scale.  Because of these two accidents, in GMSB, the MSSM
does not have a stable particle at the soft mass scale.  Provided a
hidden sector does not have such accidents, it can provide a good dark
matter candidate at the hidden soft SUSY-breaking scale.

\section{Hidden Sector Symmetries}

Thus far, we have not discussed the nature of the symmetry which
stabilizes the hidden sector dark matter particle.  This symmetry
could be a gauge, global or discrete symmetry.  Discrete symmetries
can naturally arise from the breaking of gauge symmetries; for
example, if a field in the symmetric representation of a U(N) gauge
group gets a vacuum expectation value, a $Z_2$ subgroup of the
diagonal U(1) will survive.

We can then classify the continuous symmetries that are unbroken at
the scale where the messengers are integrated out (so the mass of the
gauge boson is much lighter than the messenger mass).  The difference
between a ``gauge" and ``global" symmetry, for our purpose, is the
strength of the gauge coupling of the stabilizing symmetry ($g_s$)
with respect to the dominant gauge coupling $g_h$ of the group under
which the messengers are charged (which could, of course, be the same
symmetry).  We can think of the dark matter as being stabilized by an
unbroken ``global" symmetry if $g_s \ll g_h$ (the symmetry is truly
global if $g_s=0$).  If $g_s \sim g_h$ then we think of it as
stabilized by an unbroken gauge symmetry.

The cosmological history depends in detail on whether the stabilizing
symmetry is gauged or global, because a gauged stabilizing symmetry
implies that dark matter interacts largely through ``dark radiation."
This case is discussed in detail in~\cite{hiddensectorconstraint}.

\section{Detection}

We have not discussed the detection possibilities for this scenario;
they are discussed in detail in~\cite{wimplessdetect}, with a focus on
Yukawa couplings between dark matter and Standard Model particles via
connector particles charged under both the Standard Model and the
hidden sector.  Interestingly, since this model naturally includes
multiple hidden sectors that can each have a dark matter candidate, it
includes the possibility of multi-component dark matter, with
candidates at widely differing mass but with each composing a sizable,
${\cal O}(1)$ fraction of the relic density.  Such a scenario provides
the possibility of explaining several different experimental hints of
dark matter as discussed, for example, in~\cite{Zurek:2008qg}.


\begin{theacknowledgments}
We are grateful to the organizers of SUSY09, and thank J.~Learned,
L.~Strigari, and X.~Tata for discussions and collaboration. The work
of JLF was supported in part by NSF grant PHY--0653656.
\end{theacknowledgments}



\bibliographystyle{aipproc}   



\end{document}